\documentstyle[seceq,preprint]{jpsj} 
\def\runtitle{Impurity induced circular currents in chiral
superconductor}
\def\runauthor{Yukihiro {\sc Okuno}, 
Masashige {\sc Matsumoto} and
Manfred {\sc Sigrist}}   
\title
{Analysis of impurity-induced circular  currents for the chiral
superconductor Sr$_{2}$RuO$_{4}$ }

\author
{Yukihiro {\sc Okuno}\footnote{E-mail: okuno@yukawa.kyoto-u.ac.jp},
Masashige {\sc Matsumoto}${}^{\dagger}$  
and  Manfred {\sc Sigrist}}

\inst
{ Yukawa Institute for Theoretical Physics, Kyoto University, Kyoto 
606-8502, Japan \\
${}^{\dagger}$Department of Physics, Faculty of Science, Shizuoka 
University, 836 Oya, Shizuoka 422}

\abst
{The microscopic mechanism of circular currents induced in the
vicinity of a non-magnetic impurity is analyzed for the
(time-reversal symmetry breaking) chiral superconducting state 
$ {\bf d}({\bf k}) = \hat{{\bf z}} (k_x \pm i k_y) $ which is very
likely realized  in Sr${}_{2}$RuO${}_{4}$. 
From the analytic, not self-consistent solution of the Bogolyubov-de
Gennes equations we find two types of quasiparticle states, a bound
state and the continuum state. Through impurity scattering the
condensate transfers angular momentum to the quasiparticle states
which generate a circular current. At low temperature the continuum
part of the quasiparticle spectrum gives the main contribution. 
The non-selfconsistent solution yields a state of finite angular
momentum. The comparison with the
corresponding Ginzburg-Landau theory reveals the existence of a compensating
counterflow created by the superconducting condensate. The currents
and magnetic fields appearing around the impurity have possibly been
observed by muon spin relaxation measurements. 
}

\kword
{
Sr$_2$RuO$_4$, $p$-wave superconductor, unitary state, time-reversal
breaking state,B-dG equation,
non-magnetic impurity,spontaneous current
}

\begin{document}
\sloppy
\maketitle

\section{Introduction}
Sr${}_{2}$RuO${}_{4}$ is the first 
example of copper-free 
layered perovskite superconductor\cite{rf:Maeno}.
This material is 
structurally identical with the layered perovskite 
La${}_{2}$CuO${}_{4}$, one of the parent compounds of 
high temperature superconductors.
In spite of their structural similarity, 
there are clear differences in the electronic properties between the
two systems. The cuprates are antiferromagnetic Mott insulators in
their stoichiometric composition and turn into itinerant electron
systems with superconductivity only upon doping carriers. On the other 
hand, the normal state of 
 Sr${}_{2}$RuO${}_{4}$ displays Fermi liquid behavior with essentially
two-dimensional character and renormalized due to strong correlation
effects. Because superconductivity appears on the background of strong
electron correlation, it is unlikely that the pairing channel is
conventional s-wave type. It was early suggested that this system
might choose spin-triplet p-wave pairing symmetry instead for various
reasons \cite{rf:Rice}. First it resembles the feature of a strongly
correlated Fermi  
liquid like $^3$He which is a well-known p-wave superfluid. Second,
Sr$_2$RuO$_4$ belongs to the Ruddelsen-Popper series
Sr$_{n+1}$Ru$_n$O$_{3n+1}$ which consists almost exclusively of
ferromagnetic compounds, so that the end member Sr$_2$RuO$_4$ may be
subject to strong ferromagnetic spin fluctuations. 

Indeed there is now overwhelming evidence for the realization of
unconventional superconductivity in Sr$_2$RuO$_4$
\cite{rf:Ishida1,rf:Mackenzie} and the pairing symmetry has most likely 
the form $ {\bf d} ({\bf k}) = \hat{{\bf z}} (k_x \pm i k_y) $ analog
to the (spin-triplet) A-phase of $^3$He. The identification of the
symmetry is basically possible due to two experiments. One is the
measurement of the spin susceptibility in the superconducting state
using $^{17}$O-NMR Knight shift. According to this experiment no
change of the spin susceptibility appears with the onset of
superconductivity indicating equal spin pairing (spin triplet) 
with the moments in the basal plane \cite{rf:Ishida2}. Note that a 
strong reduction of the spin susceptibility is common to spin singlet
superconductors. Furthermore, $\mu$SR zero field relaxation rate
experiments show a pronounced increase of intrinsic magnetic field 
in the superconducting phase, suggesting that the pairing state breaks 
time reversal symmetry $ {\cal T} $ \cite{rf:Luke}. Both experiments
are fully consistent with one   
single state, $ {\bf d} ({\bf k}) = \hat{{\bf z}} (k_x \pm i k_y) $, 
among the possible pairing states.

The observation of the enhancement of the internal magnetic field is 
directly related with $ {\cal T} $-violation. In the case of the state
$ {\bf d} ({\bf k}) = \hat{{\bf z}} (k_x \pm i k_y) $ the Cooper pair
has an orbital angular momentum parallel to the $z$-axis of the
crystal. This pairing state belongs to the class of ``ferromagnetic'' ${\cal
T}$-violating superconducting states according to the symmetry
classification by Volovik and Gorkov \cite{rf:volovik}, It is
natural to expect that the presence of this angular momentum appears
in magnetic properties. In fact, however, the effect of the angular
momentum is invisible in the homogeneous superconducting phase due to
Meissner screening. It only occurs where the superconducting state is
disturbed in some way that screening effects are insufficient. This
happens for example at the surface of samples and also at domain wall
between the phases $ \hat{{\bf z}} (k_x + i k_y)  $ and $ \hat{{\bf
z}} (k_x - i k_y) $. Furthermore defects of the crystal lattice, in
particular impurities, can also lead to the appearance of the unusual magnetic 
properties of the $ {\cal T} $-violating state. It is the aim of this
paper to analyze the effect of scattering by a single impurity on the
quasiparticle states and the resulting magnetic properties. It is known that
spontaneous supercurrents are generated in the vicinity of an
impurity in such a state ~\cite{rf:Gorkov,rf:Choi,rf:Mineev}. 
This has probably been first investigated by Rainer and Vurio in the
case of the A-phase of superfluid $^3$He \cite{rf:Rainer}. 

Our analysis is based on the solution of the Bogolyubov-de Gennes
equations for the single impurity problem. In this formulation we can
show how quasiparticle states of different angular momentum around the 
impurity are coupled together and give rise to a circular current and
magnetic field. This feature can already been observed by solving the
problem on a phenomenological level with Ginzburg-Landau (GL) theory. 
Thus before giving a detailed microscopic analysis we will introduce
the problem by GL theory.

\section{Impurity problem in the Ginzburg-Landau theory}

We begin our discuss of the superconducting state around an impurity
by the discussion within the framework of GL theory. This 
allows us to recognize some of the basic feature of the
problem. The order parameter of the superconducting state has two
complex components so that we can write in general the gap function $ {\bf d}
$ as

\begin{equation}
{\bf d} ({\bf k}) = \eta_+ {\bf d}_+({\bf k}) + \eta_- {\bf d}_-({\bf
k})
\end{equation}
with $ {\bf d}_{\pm} ({\bf k}) = \hat{{\bf z}} (k_x \pm i k_y)  $. The 
free energy expansion in these two components has the form
~\cite{rf:Sigrist2,rf:Agterberg}
\begin{eqnarray}
f &=& -(|\eta_{+}|^{2}+|\eta_{-}|^{2})+\frac{1}{2}(
|\eta_{+}|^{4}+|\eta_{-}|^{4})+2|\eta_{+}|^{2}|\eta_{-}|^{2}
 \nonumber \\
&+&|{\bf D}\eta_{+}|^{2}+|{\bf D}\eta_{-}|^{2} + 
\frac{1}{2}\{ (D_- \eta_+)^* (D_+ \eta_-) + (D_- \eta_+) (D_+
\eta_-)^* \} + B^2
\end{eqnarray}
where $D_{\mu}=-i\nabla_{\mu }/\kappa-A_{\mu }$, $ D_{\pm} =D_x \pm
i D_y $ and $B={\bf \nabla} \times A$.  The free energy
density $f$ is given in units of $B^{2}{}_{c}/4\pi $ ($B_c $:
thermodynamical critical field), and lengths are measured in units of the 
penetration depth $\lambda $ and $B$ in the unit $\sqrt{2}B_{c} 
= \phi_{0}/(2\pi \lambda\xi )$, and $\kappa = \lambda /\xi $.
This form of the free energy density corresponds to the weak-coupling
approach with a single cylindrical symmetric Fermi surface.

The uniform state is immediately obtained by minimizing the free
energy with respect to the order parameter,

\begin{equation}
(\eta_+,\eta_-) = \eta_0 (1,0) \qquad {\rm or} \qquad \eta_0 (0,1)
\end{equation}
with $ |\eta_0|^2 =1 $. Either the state $ {\bf d}_{+} ({\bf k}) $ 
or $ {\bf d}_{-} ({\bf k}) $ are stabilized in the homogeneous phase. 
The gradient terms, however, show us that this needs not to be true in 
general if the superconducting phase is inhomogeneous as we will see
immediately. We will in the following assume that the homogeneous bulk 
phase is $ (\eta_+, \eta_-)  = \eta_0 (1,0) $.

Let us now add a term to the free energy introducing the effect of the
single impurity located at $ {\bf r} = 0 $, 

\begin{equation}
f_{\rm imp} = \frac{S}{2} (|\eta_+|^2 + |\eta_-|^2) \delta({\bf r})
\end{equation}
where we use coefficient $ S $ to describe the strength of the
scattering potential, noting that conventional impurity scattering is
pair breaking for unconventional superconductors, in general, leading
to a local suppression of the order parameter. 
We will analyze the behavior of the order parameter in the vicinity of 
the impurity by assuming weak distortion of the order parameter for
distances sufficiently far from the impurity. Thus we introduce two
complex functions $ u_+({\bf r}) $ and $ u_-({\bf r}) $ with
$ \eta_+ = \eta_0 + u_+({\bf r}) $ and 
$ \eta_- = u_{-}({\bf r}) $. Neglecting in addition
the effect of a finite vector potential on the order parameter we
obtain the following coupled GL equation in first order
in the impurity coupling strength.

\begin{equation} \begin{array}{l}
\displaystyle u_{+}(r)+
\eta_{0}^{2}u_{+}{}^{\ast}(r)- \frac{\nabla^2}{\kappa^2} u_{+}(r)
- \frac{\nabla_+^{2}}{2 \kappa^2} u_{-}(r)+
\eta_{0}S\delta (r)=0 \\ \\
\displaystyle u_{-}(r)- 
\frac{\nabla^2}{\kappa^2} u_{-}(r)- \frac{\nabla_-^{2}}{2
\kappa^2} u_{+}(r) =0 .
\label{GL1}
\end{array} \end{equation}
where we use $ \nabla_{\pm} = \nabla_x \pm i \nabla_y $. 
This linear equation system is easily solved by using Fourier
transformation. If we set $ \eta_0 = e^{i \chi} $, we obtain the
following solution,

\begin{equation} \begin{array}{ll}
e^{-i\chi }u_{+}(r) = & \displaystyle \int^{\infty}_{0} {\rm d}qq
\frac{\displaystyle -S(1+\frac{q^{2}}{\kappa^{2}})}{\displaystyle
2+\frac{3}{\kappa^{2}} 
q^{2}+\frac{3}{4\kappa^{4}}q^{4}}2\pi J_{0}(qr) =
g_{1}(r)  \\ & \\
e^{-i\chi }u_{-}(r) =& \displaystyle -e^{-2i\theta}\int^{\infty}_{0}
{\rm d}qq\frac{\displaystyle \frac{S}{2\kappa^{2}}q^{2}}
{\displaystyle 2+\frac{3}{\kappa^{2}}
q^{2}+\frac{3}{4\kappa^{4}}q^{4}}2\pi J_{2}(qr) =
g_{2}(r)e^{-2i\theta }.
\end{array} \end{equation}
The important point of this solution is the fact that 
$ \eta_- $-component is induced around the impurity and has a phase
winding of $ - 2 \times 2 \pi $. Any spatial variation of $ \eta_+ $
drives $ \eta_- $ which carries a phase winding.

This feature leads now to a spontaneous circular current which we
easily find by variation of the free energy with respect to the vector 
potential and the Maxwell equation $ \nabla \times {\bf B} = {\bf j}
$. Since no current is flowing in the homogeneous case we express the
current $ {\bf j} $ in first order in S,

\begin{equation} 
{\bf j} = i \eta^*_0 [ \hat{\sigma}_0 \mbox{\boldmath $ \nabla$}u_{+} +
\frac{1}{2} ( \hat{\sigma}_z + i \hat{\sigma}_x ) \mbox{\boldmath $
\nabla $}u_{-} ]  + c.c.
\end{equation}
where $ \sigma_0 $ is the $ 2 \times 2 $ unit matrix and $
\sigma_{\mu} $ the Pauli matrices. Also here we neglect the vector
potential. With the above solution we find the circular current as

\begin{equation} \begin{array}{ll}
j_{\theta}(r) & \displaystyle =  \frac{1}{\kappa}(\frac{\partial g_{2}(r)}
{\partial r}+\frac{2}{r}g_{2}(r) ) \nonumber  \\ \\
&= \displaystyle \frac{-1}{\kappa}\int_{0}^{\infty}{\rm d}q 
 \frac{2\pi \frac{S}{2\kappa^{2}}q^{4}}
 {2+\frac{3}{\kappa^{2}}q^{2}+\frac{3}{4\kappa^{4}}q^{4}}
 J_{1}(qr) \nonumber \\ \\
 &= \displaystyle -\frac{4\pi S\kappa^{2}}{\beta -\alpha }
(\beta^{\frac{3}{2}}K_{1}(\sqrt{\beta }\kappa r) -
\alpha^{\frac{3}{2}}K_{1}(\sqrt{\alpha }\kappa r))
\label{curr1}
\end{array} \end{equation}
where $\alpha=2-\frac{2}{\sqrt{3}}$, 
$\beta=2+\frac{2}{\sqrt{3}}$ and $K_{1}(x)$ is the 
modified Bessel function of first order. This expression shows that
the flow direction of the current changes as we move away from the
impurity. The current generated near the impurity is compensated again 
farther out. This kind of counter flow has been reported also by
Rainer and Vurio in the A-phase of $^3$He \cite{rf:Rainer}. 

At distances far from the impurity the current decays exponentially, $
j_{\theta} (r) =  
-  S \kappa \alpha^{5/4} \sqrt{3\pi^3 \kappa/z} \exp (-\sqrt{\alpha}
\kappa r) $ with a length scale comparable to the coherence length. 
At short distances the expression in eq.(\ref{curr1}) suggests a $
1/r$-divergence. This behavior is certainly not appropriate for short
distances. Thus, we need to introduce a
cutoff into the integral in eq.(\ref{curr1}) which leads to $
j_{\theta} (r) \propto r $ for $ r \to 0 $.

The current and counter current flow in a
narrow range around the impurity. The resulting total angular momentum 
of this flow pattern or, equivalently, the magnetic moment vanish due
to the exact canceling of the two circular currents,

\begin{equation}
{\bf M} = \frac{1}{2}\int d^{2}r {\bf r} \times {\bf j}(r) = 0 
\end{equation}

Since the canceling of the current occurs on a
length scale shorter than London penetration depth (in type II
superconductors) the Meissner screening effect does not play an
important role for the counter flow. 
Among the two length scales $ 1/ \kappa \sqrt{\beta} $ and $ 1/ \kappa
\sqrt{\alpha} $, the shorter one describes the building up of the
magnetic moment due to the order parameter distortion around the
impurity and the longer one the decay of the $ \eta_- $-component
which due to the phase winding introduces the compensating counter
flow. Note that this winding does not introduce a topological charge
like a finite flux of a vortex, since the winding order parameter
component does not exist in the bulk. As a consequence there can not
be a finite magnetic flux associated with the impurity state although
there is a local magnetic field.

\section{Bogolyubov-de Gennes Formulation}

We turn now to the description of the impurity problem in the
Bogolyubov-de Gennes (BdG) formulation in order to understand the origin of
the spontaneous currents on the level of quasiparticle states. It is
possible to perform most of the calculations analytically, if we do
not require self-consistency of the pair potential, but leave it
constant also in the vicinity of the impurity. We will discuss the
shortcomings in the approximation later. 

The uniform pair potential $ \hat{\Delta} ({\bf k}) = i (
\mbox{\boldmath $ \hat{\sigma}  $} \cdot {\bf d}({\bf  
k})) \sigma_y $ is given by 

\begin{equation}
{\bf d}(k) = \Delta _{0}(k_{x}+{\rm i}k_{y})\hat{\bf{z}} 
\label{unit}
\end{equation}

The BdG equation can be written in the following non-local form

\begin{equation} \begin{array}{l}
{\rm h_{0}}({\bf r}) u_{\rm i \sigma}({\bf r}) 
+ \sum_{\sigma'} \int d{\bf r^{'}} \hat{\Delta }_{\sigma,\sigma'}
({\bf r},{\bf r^{'}}) v_{\rm i \sigma'}({\bf r^{'}}) 
= {\rm E_{\rm i}}u_{\rm i \sigma}({\bf r}) \\
-{\rm h_{0}}({\bf r})v_{\rm i \sigma}({\bf r}) 
- \sum_{\sigma'}\int d{\bf r^{'}} \hat{\Delta^{\ast }}_{\sigma \sigma'}
({\bf r},{\bf r^{'}})u{}_{i \sigma'}({\bf r^{'}}) 
= {\rm E_{\rm i}}v{}_{\rm i \sigma}({\bf r}) \label{Bogol}
\end{array} \end{equation}
with $ u_{\rm i \sigma}({\bf r}) $ and $ v_{\rm i \sigma}({\bf
r}) $ is the particle- and hole-like part of the wavefunction. 
The part ${\rm h}_{0}$ includes besides the kinetic
energy also the potential of the non-magnetic impurity located at the origin,
\begin{equation}
{\rm h}_{0}({\bf r})
={\rm h}_{{\rm kin}}+{\rm U}\delta ({\bf r})
\end{equation}
with ${\rm h}{}_{{\rm kin}} = -\frac{\nabla^{2}}{2m} -\epsilon_{F}$ 
where $ \epsilon_{F}$ is the Fermi energy. Moreover, U denotes the 
potential strength of the impurity for which we 
assume the contact type s-wave scattering.

Because we neglect the spatial 
dependence of the gap function, in particular the suppression of the 
pair potential around the impurity, we can simplify the equation by
Fourier transformation to momentum space.
\begin{eqnarray}
\xi_{k} u_{k \sigma}+ \Delta_{k}v_{k -\sigma } 
+\frac{{\rm U}}{V} \sum_{k}u_{k \sigma } &=& \epsilon u_{k \sigma} \\
-\xi_{k}v_{k -\sigma}+ \Delta^{\ast}_{k}u_{k \sigma } 
-\frac{{\rm U}}{V} \sum_{k}v_{k -\sigma } &=& \epsilon v_{k -\sigma} 
\label{Bdguni}
\end{eqnarray}
where $\Delta_{k}=\Delta_{0}
{\rm exp}({\rm i}\phi )$ (we neglect the gap's
dependence on the magnitude  
 of the wave vector $|{\rm k}|$ 
and take the value of the 
gap at the Fermi level for simplicity.), 
$ \xi_k = k^2/2m - \epsilon_F $. 
By
introducing the variables 
${\rm I}_{0} =
({\rm U}/V) \sum_{k}u_{k \sigma } \ne 0$  
and  ${\rm I}_{0}{}^{'} =
({\rm U}/V) \sum_{k}v_{k -\sigma } \ne 0$, we can express the wave
functions as
\begin{eqnarray}
u_{k \sigma} &=& \frac{(\epsilon +\xi_{k}){\rm I}_{0}
-\Delta_{k} {\rm I}_{0}{}^{'}}
{\epsilon^{2}-\xi_{k}{}^{2}-\Delta_{0}{}^{2} } \\
v_{k -\sigma} &=& \frac{\Delta^{\ast}{}_{k}{\rm I}_{0} 
-(\epsilon -\xi_{k}){\rm I}_{0}{}^{'}}
{\epsilon^{2}-\xi_{k}{}^{2}-\Delta_{0}{}^{2} }.
\end{eqnarray}
Both $ I_0 $ and $ I'_0 $ should be determined self-consistently
leading to the equations for the energy,
\begin{equation} \begin{array}{l} \displaystyle 
{\rm I}_{0} = \frac{{\rm U}}{V} \sum_{k}u_{k \sigma}=
\frac{{\rm U}}{V} \sum_{k} \frac{(\epsilon +\xi_{k}){\rm I}_{0}
-\Delta_{k} {\rm I}_{0}{}^{'}}
{\epsilon^{2}-\xi_{k}{}^{2}-\Delta_{0}{}^{2} }
= {\rm I}_{0} \frac{{\rm U}}{V} \sum_{k} \frac{\epsilon }
{\epsilon^{2}-\xi_{k}{}^{2}-\Delta_{0}{}^{2} }  \\ \\
\displaystyle 
{\rm I}_{0}{}^{'} = \frac{{\rm U}}{V} \sum_{k}v_{k -\sigma} = 
\frac{{\rm U}}{V}\sum_{k} \frac{\Delta^{\ast}{}_{k}{\rm I}_{0} 
-(\epsilon -\xi_{k}){\rm I}_{0}{}^{'}}
{\epsilon^{2}-\xi_{k}{}^{2}-\Delta_{0}{}^{2} } = 
{\rm I}_{0}{}^{'}\frac{{\rm U}}{V} \sum_{k} \frac{ -\epsilon}
{\epsilon^{2}-\xi_{k}{}^{2}-\Delta_{0}{}^{2} } \\ 
\label{bound1}
\end{array} \end{equation}
where for final form we used the fact that the angular integral over the gap function 
$\Delta_{k}$ vanishes and that close to the Fermi surface the
assumption of electron-hole symmetry is satisfied, i.e. the normal
state density of state is constant. Actually, there is no qualitative
change if we include particle-hole asymmetry.

We immediately see that the equations for $ {\rm I}_0 $ 
and $ {\rm I}{}^{'}_0 $ are
decoupled and give rise to different solutions. The states associated
with $ {\rm I}_0 \neq 0 $ can be considered as particle-like while the ones for
$ {\rm I}{}^{'}_0 \neq 0 $ are hole-like. Within both sectors there are two
types of solutions, discrete energy states and states belonging to a
continuous spectrum (Fig.2). The former consist of one midgap bound state in
both sectors and an electron-(hole)-like state above (below) the band
top (bottom) which correspond to antibound states (for repulsive
impurity scattering). In the following we will neglect these anti-bound 
state, since they are very far from the Fermi level and will not have
much influence on the properties we are interested in.
It is useful to separate the discussion of the midgap bound states and 
the states of the continuum. 

\subsection{Midgap bound states}

From Fig.2 we recognize that there is one discrete midgap state for
both case 
$ {\rm I}_{0} \neq 0 $ and $ {\rm I}^{'}_0 \neq 0 $ 
with the energies
\begin{equation}
\epsilon_\mp = \mp \frac{\Delta_0}{\sqrt{1 + c^2}}
\end{equation}
respectively where $ c = \pi N_0 {\rm U} $ and $ N_0 $ is the density
of states at the Fermi level. Note, that for strong scattering ($ {\rm 
U} \to \infty $) the bound state energy are both zero and for weak
scattering ($ {\rm U} \to 0 $) the bound states are located close to
the gap edge $ \pm \Delta_0 $. 
For the case particle-like bound state 
($ {\rm I}{}_0 \neq 0 $) we obtain the wave function 

\begin{equation} \begin{array}{l} \displaystyle
u^{(-)}_\sigma ({\bf r}) = \frac{{\rm I}_{0}}{2 \pi} \int^{\infty}_{0}
 {\rm d}kk\frac{\epsilon_- + \xi_{k}}
{\epsilon_{-}{}^{2}-\xi_{k}{}^{2}-\Delta_{0}{}^{2}} J_{0}(kr)
\approx -\frac{2{\rm I}_{0}{\rm N}_{0}\epsilon_{-}}
{\sqrt{\Delta_{0}{}^{2}-\epsilon_{-}{}^{2}}}f_{1}(k_{F}r)
\\ \\  \displaystyle
v^{(-)}_{-\sigma}({\bf r}) = \frac{{\rm I}_{0}}{2 \pi} \int^{\infty}_{0}
 {\rm d}kk\frac{\Delta_{0}}
{\epsilon_{-}{}^{2}-\xi_{k}{}^{2}-\Delta_{0}^{2}}
{\rm i} J_{1}(kr){\rm e}^{-{\rm i}\theta}
\approx  -\frac{2{\rm I}_{0}{\rm N}_{0}\Delta_{0}}
{\sqrt{\Delta_{0}{}^{2}-\epsilon_{-}{}^{2}}}
{\rm i}f_{2}(k_{F}r){\rm e}^{-{\rm i}\theta} \\
\end{array} \label{bsolu1}
\end{equation}
and for the hole-like bound state ($ {\rm I}{}^{'}{}_{0} \neq 0 $)

\begin{equation} \begin{array}{l} \displaystyle
u^{(+)}_{\sigma}({\bf r}) = - \frac{{\rm I}^{'}_{0}}
{2 \pi} \int {\rm d}kk\frac{\Delta_{0}{}^{\ast}}
{\epsilon_{+}{}^{2}-\xi_{k}{}^{2}-\Delta_{0}^{2}}
{\rm i} J_{1}(kr){\rm e}^{{\rm i}\theta}
\approx \frac{2{\rm I}'_{0}{\rm N}_{0}\Delta_{0}}
{\sqrt{\Delta_{0}{}^{2}-\epsilon_{+}{}^{2}}}
{\rm i}f_{2}(k_{F}r){\rm e}^{{\rm i}\theta} \\ \\ \displaystyle
v^{(+)}_{-\sigma}({\bf r}) = - \frac{{\rm I}'_{0}}{2 \pi}
 \int {\rm d}kk\frac{\epsilon_{+} - \xi_{k}}
{\epsilon_{+}{}^{2}-\xi_{k}{}^{2}-\Delta_{0}{}^{2}} J_{0}(kr)
\approx \frac{2{\rm I}_{0}{}^{'}{\rm N}_{0}\epsilon_{+}}
{\sqrt{\Delta_{0}{}^{2}-\epsilon_{+}{}^{2}}} 
f_{1}(k_{F}r) \\ 
\label{bsolu2}
\end{array} \end{equation}
where $k_{F}$ is the 
Fermi wave number, $\theta$ is the angle of position 
vector ${\bf r}$ and $J_{0}(kr)$, $J_{1}(kr)$ is the Bessel function 
of 0th and 1st order, respectively. The form of the function 
$f_{1}(k_{F}r)$ and $f_{2}(k_{F}r)$ for  
large distances $(r>>1/k_{F})$ is approximatively given by
\begin{equation} \begin{array}{l} \displaystyle
f_{1}(k_{F}r) \approx \frac{\pi}{2}\sqrt{\frac{2}{\pi k_{F}r}}
{\rm cos}(k_{F}r-\frac{\pi}{4})
{\rm e}^{-\frac{\sqrt{\Delta_{0}{}^{2}-\epsilon^{2} }r}
{v_{F}}r} 
\\ \\ \displaystyle
f_{2}(k_{F}r) \approx \frac{\pi}{2}\sqrt{\frac{2}{\pi k_{F}r}}
{\rm cos}(k_{F}r-\frac{3\pi}{4})
{\rm e}^{-\frac{\sqrt{\Delta_{0}{}^{2}-\epsilon^{2} }}
{v_{F}}r}.
\\
\end{array} \end{equation}
where $\epsilon=\Delta_{0}  /(\sqrt{1+{\rm c}^{2}})$ and 
$v_{F}$ is Fermi velocity. \\ 
The variables 
${\rm I}_{0}$ and ${\rm I}_{0}{}^{'}$
are determined by the normalization condition,
\begin{equation}
\sum_{k}(u_{k}{}^{2}+v_{k}{}^{2})=1
\end{equation}
leading to ${\rm I}_{0}{}^{2}={\rm I}_{0}{}^{'2}=
\Delta_{0}{\rm c}^{3}/(\pi{\rm N}_{0}(1+{\rm
c}^{2})^{\frac{3}{2}})$. \\
The solution in eq.(\ref{bsolu1}) and (\ref{bsolu2})  
are very similar to the bound state 
discussed in connection with  
magnetic impurities in  conventional 
superconductors~\cite{rf:Yu,rf:Shiba} and was noticed by Buchholtz 
and Zwicknagel~\cite{rf:Buchholtz}.  
However, the important 
difference lies in the angular momentum structure of the above bound
states. We can easily see that the for the electron-like state the
particle wave functions couples with the hole wave function of an
angular momentum reduced by 1. For the hole-like states it is just
the opposite way around. This property is responsible for the fact that
these states can carry a circular current. In terms of the
wavefunctions $ u $ and $ v $ the current is expressed as,

\begin{eqnarray}
j_{B}({\bf r}) &=& \frac{{\rm e}}{2m{\rm i}}\sum_{i,\sigma}
[f(\epsilon_{i})u^{\ast}{}_{i \sigma}({\bf r})\nabla
u_{i \sigma}({\bf r})+(1-f(\epsilon_{i}))v_{i \sigma}({\bf r})
\nabla v^{\ast}{}_{i \sigma}({\bf r}) -c.c] \nonumber \\
&=& \hat{\bf {\rm e}}_{\theta}\frac{8{\rm e}\Delta_{0}
\pi{\rm N}_{0}{\rm c}}{m\sqrt{1+c^{2}}}f(\epsilon )
\frac{f_{2}(k_{F}r)^{2}}{r}
\label{current1}
\end{eqnarray}
where $\hat{\bf {\rm e}}_{\theta}$ is the unit vector of the 
$\theta$ component, $\epsilon = \Delta_{0}/\sqrt{1+{\rm c}^2 }$, 
and  $f(\epsilon)$ is 
the Fermi distribution function.
We immediately realize that this current disappears for
low-temperatures exponentially, $ \propto \exp(- \epsilon/ k_B T) $,
if U is positive (repulsive). 
Basically no spontaneous current is caused by the bound
states if consider the electron states, since the electron state
corresponds to an anti-bound state due to the repulsive potential. On
the other hand, considering the situation from the hole point of view
the potential is attractive and the hole in a real bound state such
that it carries circular current. We will discuss 
this point using the e-h transformation at the end of this section.

\subsection{Contribution of the continuous spectrum}

We turn now to the discussion of the continuous part of the
quasiparticle spectrum. 
Analyzing Fig.2 we immediately see that these states are
relevant for the formation of circular currents as they are occupied
at low temperature. In order to discuss their contribution
to the current we have to go beyond the discussion we used for the
midgap states. A possible formulation can be based on the solution of
the Lippmann-Schwinger equation for the scattering problem. Let us
start with the construction of the out-going wave as

\begin{eqnarray}
\psi^{(+)}{}_{k}({\bf r}) &=& \phi {}_{k}({\bf r}) + 
\int {\rm d}y \hat{G}^{0}{}_{k}({\bf r},{\bf r^{'}})
{\rm U} ({\bf r^{'}})\tau_{3} \psi^{(+)}{}_{k}({\bf r^{'}}) 
\nonumber \\
&=& \phi{}_{k}({\bf r}) +
\hat{G}^{0}{}_{k}({\bf r},0)
\frac{{\rm U} \tau_{3}}
{1-\hat{G}^{0}{}_{k}(0,0){\rm U}\tau_{3} }
\phi {}_{k}(0)
\label{Lippmann}
\end{eqnarray} 
where $\tau_{i}$ denotes the i-th Pauli matrix and 
the function $ {\phi }(x) $ is the solution 
of the B-dG equation without impurity given by
 
\begin{equation}
\phi{}_{k}({\bf r}) =
\left (
\begin{array}{c}
u^{(0)}{}_{k}({\bf r}) \\
v^{(0)}{}_{k}({\bf r}) 
\end{array}
\right )
= \frac{1}{
\sqrt{2V E_{k}(E_{k}+\xi_{k})}} 
\left (
\begin{array}{c}
E_{k}+\xi_{k} \\
\Delta_{k}{}^{\ast} 
\end{array}
\right ) 
{\rm e}^{{\rm i}{\bf k} \cdot {\bf r}} 
\end{equation}
for the case $ {\rm I}{}_{0} \neq 0 $, 
${\rm I}'{}_0 = 0 $ and

\begin{equation}
\phi{}_{k}({\rm r}) =
\left (
\begin{array}{c}
u^{(0)}{}_{k}({\bf r}) \\
v^{(0)}{}_{k}({\bf r}) 
\end{array}
\right )
= \frac{1}{
\sqrt{2V E_{k}(E_{k}-\xi_{k})}} 
\left (
\begin{array}{c}
\Delta_{k} \\
E_{k}-\xi_{k}  
\end{array}
\right ) 
{\rm e}^{{\rm i}{\bf k} \cdot {\bf r}} 
\end{equation}
for $ {\rm I}_0 = 0 $, 
${\rm I}'{}_0 \neq 0 $. Note that each momentum $k$
corresponds to two values of the energy,
$E_{k}=\pm \sqrt{\xi_{k}{}^{2}+\Delta_{0}{}^{2}}$. The Green function
entering eq.(\ref{Lippmann}) is constructed from these solutions 
\begin{eqnarray}
\hat{G}^{0}{}_{k}({\bf r},{\bf r^{'}}) &=& 
\sum_{k^{'}} \frac{\phi{}_{k^{'}} ({\bf r}) 
\phi{}^{\dagger }{}_{k^{'}} ({\bf r^{'}})}
{E_{k}+{\rm i}\delta -E_{k^{'}}} \nonumber \\
&=& 
\frac{1}{V}\sum_{k^{'}} 
\frac{
\left (
\begin{array}{cc}
\frac{E_{k^{'}}+\xi_{k^{'}}}{2E_{k^{'}}}
&  \frac{\Delta_{k^{'}}}{2E_{k^{'}}} \\
\frac{\Delta_{k^{'}}^{\ast}}{2E_{k^{'}}} &
\frac{E_{k^{'}}-\xi_{k^{'}}}{2E_{k^{'}}}
\end{array}
\right )
}
{E_{k}+{\rm i}\delta -E_{k^{'}}}
{\rm e}^{{\rm i}{\bf k^{'}} \cdot ({\bf r}-{\bf r^{'}})}
  \nonumber \\
&=& G^{0}{}_{k,0}(r)\tau_{0} + G^{0}{}_{k,3}(r)\tau_{3}
+ G^{0}_{k,\theta}(r)
\left (
\begin{array}{cc}
0 & {\rm e}^{\rm i\theta} \\
{\rm e}^{-{\rm i}\theta} & 0 
\end{array}
\right ) 
\label{GrenR}
\end{eqnarray}
where $r=|{\bf r}-{\bf r^{'}}|$. The Green functions  
$G^{0}{}_{k,0}(r)$, $G^{0}{}_{k,3}(r)$
and $G^{0}{}_{k,\theta}(r)$ can be expressed analytically, 
\begin{eqnarray}
G^{0}{}_{k,0}(r) &=& -\frac{2E_k\cdot {\rm sgn}E_k{\rm N}_{0}}
{\sqrt{E^2_k-\Delta_{0}{}^{2}}}
\frac{\pi}{2}{\rm i}
[ {\rm H}_{0}{}^{(1)}(k_{{\rm sgn} E_k}r) 
+ {\rm H}_{0}{}^{(2)}(k_{{\rm -sgn} E_k}r) ] 
\nonumber \\
G^{0}{}_{k,3}(r) &=& -\frac{{\rm N_{0}}}{V}
\frac{\pi}{2}{\rm i} 
[ {\rm H}_{0}{}^{(1)}(k_{{\rm sgn} E_k}r) 
- {\rm H}_{0}{}^{(2)}(k_{{\rm -sgn} E_k}r) ] \nonumber \\
G^{0}{}_{k,\theta}(r) &=& 
-\frac{{\rm i}{\rm N}_{0}\Delta_{0}}
{ \sqrt{E^2_k-\Delta_{0}{}^{2}}}
\int dkk{\rm J}_{1}(kr) \{ 
\frac{1}{k^{2}-k_{+}{}^{2}-{\rm i}\delta{\rm sgn}E_k}
-\frac{1}{k^{2}-k_{-}{}^{2}+{\rm i}\delta{\rm sgn}E_k} \} \nonumber \\
&=& \frac{{\rm N}_{0}\Delta_{0}}
{\sqrt{E{}^{2}{}_{k}-\Delta_{0}{}^{2}}}{\rm sgn}E_{k} 
\frac{\pi}{2}[{\rm J}_{1}(k_{+}r)+{\rm J}_{1}(k_{-}r) ] \nonumber \\
&-& \frac{{\rm i}{\rm N}_{0}\Delta_{0}}
{ \sqrt{E^2_k-\Delta_{0}{}^{2}}} 
{\rm P}\int dkk [ \frac{1}{k^{2}-k_{+}{}^{2}}
-\frac{1}{k^{2}-k_{-}{}^{2}} ]{\rm J}_{1}(kr)  \nonumber \\
& \equiv & g_{1,k}(r)+{\rm i}g_{2,k}(r) \label{Gdecom}
\end{eqnarray}
where ${\rm J}_{1}(x)$ is the first kind  Bessel function 
and ${\rm H}_{0}{}^{(1)}(x)$ and 
${\rm H}_{0}{}^{(2)}(x)$ are Hankel functions. 
Moreover, the momenta $ k_{\pm} $ are given by
\begin{eqnarray}
k_{\pm} &=& k_{F}\sqrt{1\pm \frac{
\sqrt{E^2_k-\Delta_{0}^{2}}}{\epsilon_{F}}} \approx k_F \pm
\frac{\sqrt{E^2_k - \Delta^2_0}}{v_F}. 
\end{eqnarray}
where $ v_F = k_F / m $ is the Fermi velocity. The two momenta $ k_+ $ 
and $ k_- $ correspond to the more particle-like and more hole-like
sections of the quasiparticle spectrum. The appearance of these two
states together is particular feature of the particle-hole mixing in
the superconducting state and is related with the Andreev reflection. 
The last of these three Green
functions will be important for the discussion of the circular
current. 

In (\ref{Gdecom} )
we decompose  $G^{0}{}_{k,\theta}(r)$ as $g_{1,k}(r)$ and 
$g_{2,k}(r)$ as it will be convenient for the later discussion. 
It is easy to see that $g_{2,k}(r)$ is much smaller than $g_{1,k}(r)$ 
for the states with $ |\sqrt{E^2_k-\Delta_{0}}| \ll \epsilon_{F}$. 
Since these states will be dominant in the contribution to the
currents, in particular, due to the enhanced density of states for
energies just above the gap, we may safely neglect $g_{2,k}(r)$. 

The above Green function can now be used to define the T-matrix,

\begin{equation}
\hat{T}(\epsilon) = \frac{{\rm U}\tau_{3}}
{1-\hat{G}^{0}_{k}(0,0){\rm U}\tau_{3} } 
\equiv T_{0}\tau_{0}+T_{3}\tau_{3}
\end{equation}
where the two components are given as,
\begin{eqnarray}
T_{0} &=& -\frac{{\rm i}\pi {\rm U}^{2}\rho (E_k)}
{1+(\pi\rho (E_k){\rm U})^{2}} \\
T_{3} &=& \frac{{\rm U}}{1+(\pi\rho (E_k){\rm U})^{2}}
\end{eqnarray}
where $\rho (E_k)\equiv -\frac{1}{\pi}{\rm Im}G^{0}{}_{k,0}(0)$. 
Note that $T_{0}$ is purely imaginary, while $T_{3}$ is 
real and vanishes in the unitary limit. This just reflects the fact that 
the scattering phase shift is $\frac{\pi}{2}$ in this limit. 

Let us now rewrite the outgoing wave solution of the Lippmann-Schwinger 
equation. We restrict to the case where 
$ {\rm I}{}_{0} =0 $ and $ {\rm I'}{}_{0} \neq 0 $ 
since the other case gives finally the same contribution to the
current. 

\begin{equation} \begin{array}{ll}
\psi_{k}{}^{(+)}({\bf r}) = & \displaystyle \left ( \begin{array}{c}u_{k}({\bf r}) \\
v_{k}({\bf r})  \end{array}\right ) 
 = \frac{1}{\sqrt{2V E_{k}(E_{k}-\xi_{k})}} 
\left (
\begin{array}{c}
\Delta_{k} \\
E_{k}-\xi_{k}  
\end{array}
\right ) {\rm e}^{{\rm i}{\bf k} \cdot {\bf r} } \\ & \\ & \displaystyle
\qquad + \frac{1}{\sqrt{2V E_{k}(E_{k}-\xi_{k})}} 
\left (
\begin{array}{c}
(S_{0}(r)+S_{3}(r))\Delta_{k} \\
(S_{0}(r)-S_{3}(r))(E_{k}-\xi_{k})  
\end{array}
\right ) \\ & \\ & \displaystyle
\qquad + 
\frac{G^{0}{}_{k,\theta}(r) }{\sqrt{2V E_{k}(E_{k}-\xi_{k})}}
\left (
\begin{array}{c}
(T_{0}-T_{3})(E_{k}-\xi_{k})
{\rm e}^{{\rm i}\theta } \\
(T_{0}+T_{3})\Delta_{k}{\rm e}^{-{\rm i}\theta } \label{conso}
\end{array}
\right )
\label{sol}
\end{array} \end{equation}
with $S_{0}(r) = G^{0}{}_{k,0}(r)T_{0}+G^{0}{}_{k,3}(r)T_{3}$ and 
$S_{3}(r) = G^{0}{}_{k,0}(r)T_{3}+G^{0}{}_{k,3}(r)T_{0}$. 
We use now this solution to calculate the current according to
eq.(\ref{current1}). Also in the present case only the angular
component is finite which we decompose into two part $ j_{\theta} (r)
= j_1 (r) + j_2 (r) $. The product of the third term in eq.(\ref{sol}) 
and its derivative leads to

\begin{equation} \begin{array}{ll}
j_{1}(r) & \displaystyle = \frac{{\rm e}}{m V} 
\sum_{k,E_{k} > 0}  \frac{|G{}^{\theta}{}_{k,0}(r)|^{2} }{r} 
(|T_{0}|^{2}+|T_{3}|^{2}) \\ & \\ & \displaystyle
\approx \frac{{\rm e}}{m } 
\int_{E_k > 0} \frac{{\rm d}k k}{2 \pi} \frac{\pi^2 N^2_0 \Delta^2_0}{4(E^2_k -
\Delta^2_0)}  \frac{({\rm J}_1 (k_+ r)+{\rm J}_1 (k_- r))^2}{r}
(|T_{0}|^{2}+|T_{3}|^{2})
\label{cur1}
\end{array} \end{equation}  
where we used that $ k_{\pm} \approx k_F $ and $ g_2 \ll g_1 $ 
in eq.(\ref{Gdecom}). 
Next we consider the contribution coming from the product of the first 
and the third term in eq.(\ref{sol}), again neglecting the
contribution of $ g_2 $,

\begin{equation} \begin{array}{ll}
j_{2}(r) & \displaystyle = \frac{{\rm e}}{m{\rm i}}[ 
\int \frac{{\rm d}kk}{2\pi} \frac{g_{1,k}(r)}{E_{k}}T_{0}\Delta_{0}
\frac{J_{1}(kr)}{r} -c. c. ] \\ & \\
&  \displaystyle \approx \frac{{\rm e}}{m{\rm i}} \int_{E_k > 0}
[\frac{{\rm d}kk}{2\pi} \frac{\pi N_0 \Delta^2_0 T_0 }{2\sqrt{E^2_k -
\Delta^2_0}} \frac{({\rm J}_1(k_+ r) + {\rm J}_1 (k_- r)){\rm J}_1 (k r))}{E_k r} - c.c. ] \\
\label{cur2}
\end{array} \end{equation}
Note that the expression of the current (\ref{cur1},\ref{cur2}) the 
the Fermi distribution function does not appear and the 
only temperature dependence enters only via $\Delta_{0}$.
In the unitary limit ($ {\rm U} \to \infty $) $ T_3 $ vanishes and the 
total circular current which comes from the continue parts  
reduces to rather simple form
\begin{equation}
j_{\theta}(r) =  \frac{{\rm e}}{mr} \int \frac{{\rm d}k k}{2 \pi}
\frac{\Delta^2_0}{E^2_k} \left[ \frac{1}{16} \{ {\rm J}_1(k_+ r) +
{\rm J}_1 (k_- r)\}^2 - \frac{1}{2} ({\rm J}_1(k_+ r) + {\rm J}_1 (k_-
r)) {\rm J}_1 (k r) \right]
\label{finalcurrent}
\end{equation}
The main contribution of this integral comes from the states at the
gap edge with the wave vector close to $ k_F $. 
In this form it is easy to analyze the long-distance behavior of 
$ j_{\theta}(r) $, i.e. $ r \gg 1 / k_F $. 

\begin{equation}
j_{\theta} (r) \approx  - \frac{7 {\rm e} \Delta_0}
{8 \pi v_{F}r^2 } [1 + e^{- 2
\frac{\Delta_0}{v_F }r}]
 \cos^2(k_F r - \frac{3}{4}\pi) \label{concur}
\end{equation}
where $v_{F}$ is Fermi velocity and 
both components of the current show the same $ r $-dependence. 
We observe two distinct
contributions showing Friedel-type oscillations, one is exponentially
fast decaying and the other follows the power-law $ 1/r^2 $ (Fig.3). 
The latter is a particular result of the fact that the current is
carried by the quasiparticles belonging to the continuum spectrum. 
This term does not fit well with the solution of the GL
theory where we found an exponential decay behavior including also the 
flow of counter currents. While the power-law behavior may be appropriate
in an intermediate length regime the GL treatment
suggests that it should not apply for long distances
\cite{rf:Rainer}. Definitely the long-range extension of the circular
current should be inhibited by the Meissner effect which yields
screening counter currents living on the length scale $ \lambda $, the 
London penetration depth. This effect is not included in our
discussion here, as it was omitted in the analysis of the
GL theory. A further more serious omission is, however, the 
self-consistency of the pair potential $ \Delta $. In the GL theory
the reaction of the superconductor to the impurity was the admixture
of superconducting phase with opposite chirality, i.e. locally the
state $ {\bf d} ({\bf k}) = \hat{{\bf z}} (k_x - i k_y) $ is admixed
to the bulk state 
$ {\bf d} ({\bf k}) = \hat{{\bf z}} (k_x + i k_y)$. \\

\subsection{Electron-hole transformation}

As mentioned above the repulsive impurity potential problem considered 
from the electron-hole converted point of view corresponds to a
problem with attractive potential. This can be easily seen by applying 
electron-hole transformation which means that we replace

\begin{equation}
{\rm h}{}_{0,h}({\bf r})=-{\rm h}{}_{0,e}({\bf r}), 
\hspace{0.5cm} \hat{\Delta_{h}}({\bf r},{\bf r^{'}})=
-\hat{\Delta_{e}}{}^{\ast}({\bf r},{\bf r^{'}}) \hspace{0.5cm}
({\rm u_{h,i}},{\rm v_{h,i}})=({\rm v_{e,i}},{\rm u_{e,i}})
\end{equation}
Here the subscripts $e$ and $h$ denote the problem in the electron or 
hole perspective, respectively. It is important to notice that the
transformation of the gap is actually an inversion of the Cooper pair
angular momentum, ${\bf d}({\bf k}) = \hat{{\bf z}}(k_{x}+{\rm i}k_{y}) 
\rightarrow \hat{{\bf z}}(k_{x}-{\rm i}k_{y})$.
The solution for the so generated attractive impurity problem gives
the same current as the original repulsive case.

\begin{eqnarray}
j_{h}({\bf r}) &=& \frac{{\rm e}}{2m{\rm i}}\sum_{i,\sigma}
[f(\epsilon_{i}){\rm u}^{\ast}_{h i,\sigma}\nabla {\rm u}_{h i,\sigma}
+(1-f(\epsilon_{i})){\rm v}_{h i,\sigma}\nabla 
{\rm v}^{\ast}_{h i,\sigma}-{\rm c.c}] \nonumber \\
&=& j_{e}({\bf r}) 
\end{eqnarray}
The main difference 
between the repulsive and the attractive one is the 
contribution of the bound state  solution to the spontaneous current
at the zero temperature.
In the hole perspective, using the bound state 
solutions (\ref{bsolu1}) and (\ref{bsolu2}), 
the contribution of the bound state is given as 
\begin{eqnarray}
j_{B}({\bf r}) &=& -
\hat{\bf {\rm e}}_{\theta}\frac{8{\rm e}\Delta_{0}
\pi{\rm N}_{0}{\rm c}}{m\sqrt{1+c^{2}}}(1-f(\epsilon ))
\frac{f_{2}(k_{F}r)^{2}}{r}. \label{hbcur}
\end{eqnarray} 
which is non-vanishing. The continuum part compensates for this
change, but has basically the same structure as eq.(\ref{concur}) with 
slightly different coefficients. Therefore, the sign of the impurity potential
does not change the spontaneous circular current, but only the view point.

\section{Discussion}

We have considered a superconducting state whose Cooper pairs have a
finite angular momentum, a chiral state. Clearly the total angular
momentum of the system should be zero in the ground state or in
thermodynamic equilibrium in the absence of an external magnetic
field. Nevertheless, in the vicinity of an impurity the intrinsic
angular momentum becomes visible in form of a circular current. 
In the Bogolyubov-de Gennes formulation we have seen that these
circular currents originate mainly from quasiparticle scattering
states belonging to the continuous part of the spectrum, while the
quasi particle bound state at the impurity plays in this respect a
minor role. This calculation has been done without taking
self-consistence into account. Thus the order parameter is uniform and 
does not get a feedback from the modified quasi particle spectrum. 
As a consequence the quasiparticle states, a hybridization of electron- 
and hole-like states of different angular momentum yield a 
finite total angular momentum if we do not include the edge of the
system. The edge introduces chiral quasiparticle modes and a
spontaneous surface current \cite{rf:Matsumoto}. In our solution of the
B-dG problem the impurity potential leads to a shift of the electron
concentration and some amount of charge is transfered to the edge
modes which have zero-energy states. This in turn leads to a change of 
the edge currents and to a compensating change of angular momentum. 
In our solution this is connected with the circular current which
decays like $ r^{-2} $. Thus the total angular momentum of a finite
system is not changed even in the non-self-consistent solution of the
B-dG equations. 

On the other hand, the GL formulation leads to a canceling of the
angular momentum basically within the short length scale $ \xi $. Thus 
the feedback of the superconducting condensate leads to a strong
screening and the impurity and edge problem are decoupled from each
other. Thus the above connection has to be considered as artificial.
This type of screening was recently also discussed in detail by Kusama 
and Ohashi~\cite{rf:Kusama} in connection with Bloch's theorem.

While it is difficult to include this screening effect in an analytical
way into the B-dG  formulation, we can see this property very
conveniently in the GL theory where around the impurity the $ \eta_- $
component appears in 
addition to the bulk phase corresponding to $ \eta_+ $. This admixed
order parameter has a phase winding and introduces vorticity yielding
a counter flowing current. Since this order parameter decays towards
the bulk the vorticity does not constitute a real vortex and no
magnetic flux exists. The $ \eta_- $ component disappears over a
length comparable to the coherence length and with it counter currents 
which within this length completely compensate the angular momentum
introduced by the impurity scattering states. Overall seen there is no 
net magnetic moment or magnetic flux. Consequently, the circular currents 
at the impurity are not easy to observe. Magnetometers would need a resolution
higher than coherence length. One such probe is provided by spin
polarized muons in zero magnetic field and has recently successfully 
lead to the observation of intrinsic magnetism in the superconducting
state of Sr$_2$RuO$_4$ \cite{rf:Luke}. 

Recently, Goryo and Ishikawa have proposed an experiment which has some 
similarity to the impurity problem treated here \cite{rf:Goryo}. Their
analysis is based on an topological argument that the chiral p-wave
state introduces an effective Chern-Simons term into the Ginzburg-Landau
theory, which couples the vector potential with the scalar potential. 
A charge inserted into the superconductor should  generate a circular
current similar to a Hall current. There is clear analogies between 
the two studies in various aspects and it is certain worthwhile to
investigate the connections more carefully. In particular, the argument
about the vanishing angular momentum should apply for the Hall current of the
Chern-Simon term as well in a finite system,
since the external magnetic field is absent. 

The physics associated with the chiral quasiparticle states around the
impurity as well as at the surface can be discussed in terms of 
Andreev bound states as they were investigated first by Buchholtz and
Zwicknagel \cite{rf:Buchholtz} for $p$-wave states and in a related
form for $d$-wave states by Hu \cite{rf:Hu} and many others later.
In the case of $ d$-wave superconductivity, however, the topological
argument discussed in Ref.\cite{rf:Goryo} does not apply and the
zero-energy states at surfaces and impurities are not
chiral. Nevertheless, they have other important implications \cite{rf:Kashiwaya}.

The topological aspect is connected with the presence of an internal
angular momentum of the Cooper pair of the type $p_{x} \pm {\rm
i}p_{y}$, $ L_z = \pm 1 $. Thus, similar physics is expected in other
$ {\cal T}$-violating superconducting states of the same or similar
topology (angular momentum). To these belong the two $ d $-wave states 
$ d_{x^2-y^2} \pm id_{xy} $ with $ L_z = \pm 2 $ and $ d_{zx} \pm i
d_{zy} $ with $ L_z = \pm 1 $. On the other hand, the $ {\cal
T}$-violating state $ d_{x^2 - y^2} \pm i s $ often discussed in
connection with high-temperature superconductors does not belong to
this class as it does not have an angular momentum. Consequently, we
do not find a simple circular current around impurities in this case,
in contrast to all other case mentioned above. The current pattern in
this case is more complicated and will be discussed elsewhere.

\section{Acknowledgement}

We would like to thank A.Furusaki, R. Heeb and T. Tamaribuchi for helpful 
discussions.  This work was supported by Grant-in-Aid 
for Scientific Reserch from Ministry of Education. 
This work is supported by the Grant-in-Aid for 
Scientific Research (10740169) and (10640341), and
one of the authors (Y.O) has been supported by a C.O.E. fellowship 
from the Ministry of Education, Science, Sports and Culture of Japan.

\newpage
Figure caption
\begin{itemize}         
 \item Fig.1 Circular current around the impurity derived from the GL
             theory. We take the value $\kappa =7$, 
             and the length is renormalized by the penetration 
             depth $\lambda$ and the current 
             by $4\pi S\kappa^{2}$. The expression of  
             (\ref{curr1}) (solid line) 
             is not valid in short distance 
             and the current should be 0 for $r \rightarrow 0$.
             We omit this in our graph as the length scales are
             rather different. 

 \item Fig.2 Schematic view of the 
             solutions of the B-dG equation. 
             Circles ( diamond ) correspond to the continuous 
             ( bound ) states. The solutions in the  
             first and the third (the second and the forth ) 
             quadrants are particle ( hole ) like 
             states.
 \item Fig.3 Current around the impurity from eq.(\ref{finalcurrent})
             . We renormalize the energy by $\Delta_{0}$, and the
             length by the coherence length $\xi = v_{F}/\Delta_{0}$,
             and the current by $e/(2m)$. 
\end{itemize}
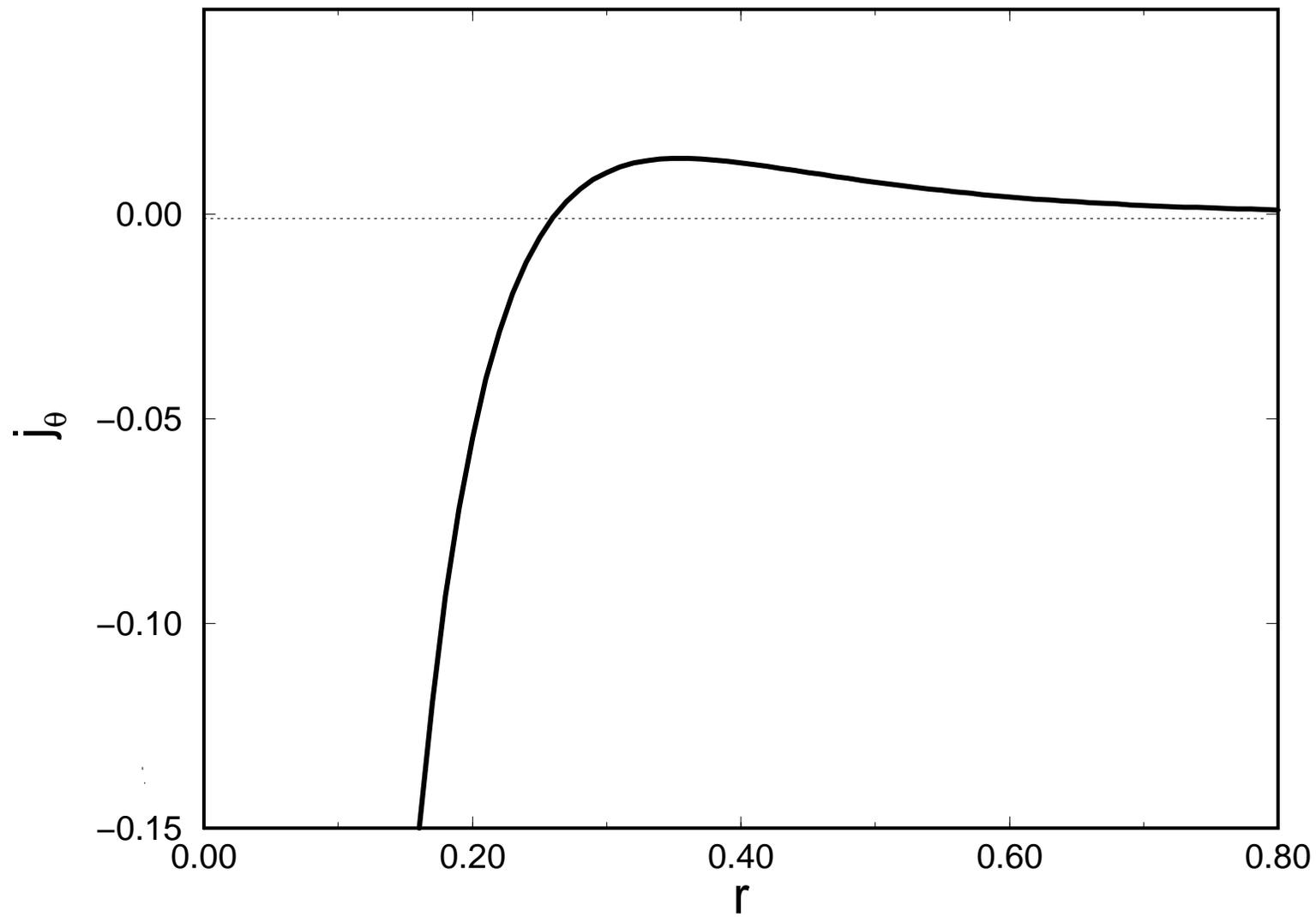
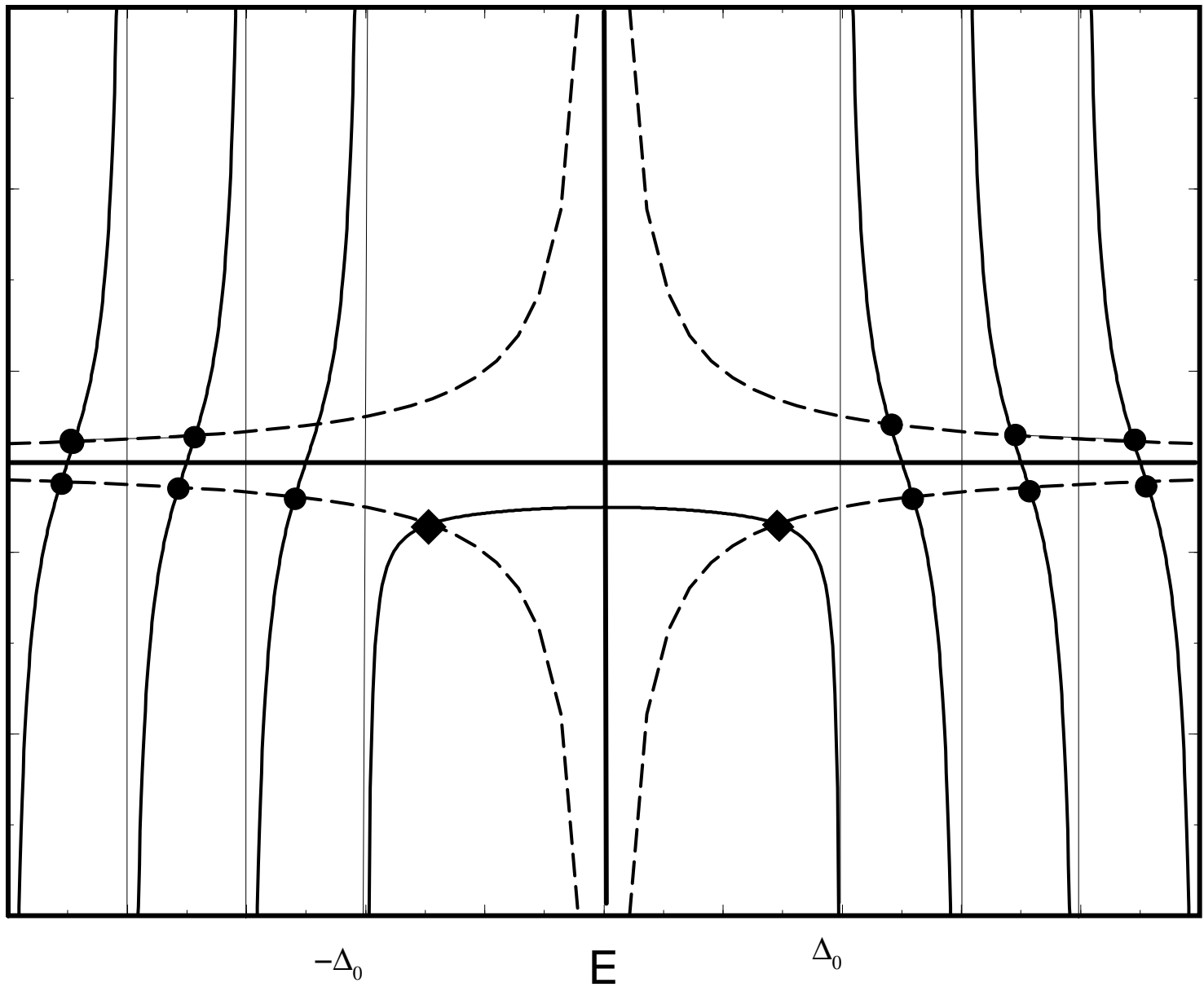
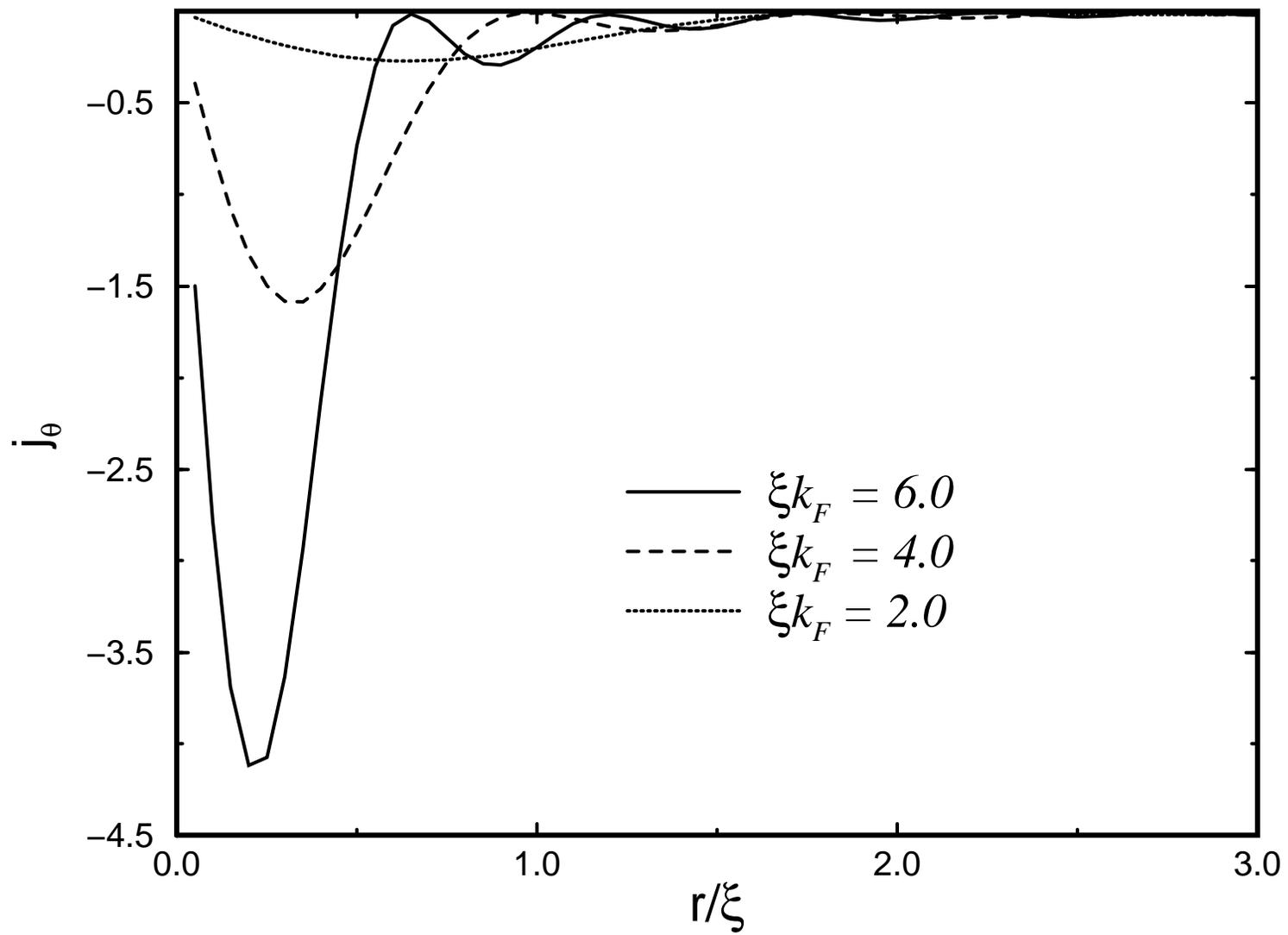
\end{document}